\documentclass[pra,superscriptaddress,reprint]{revtex4-2}
\usepackage[utf8]{inputenc}
\usepackage[fleqn]{amsmath}
\usepackage{amsfonts}
\usepackage{amssymb}
\usepackage{graphicx}
\usepackage{epstopdf}
\usepackage{subcaption}
\usepackage[T1]{fontenc}
\usepackage{color} 
\usepackage{url}
\usepackage{float}
\usepackage[hidelinks]{hyperref}
\usepackage[toc,page,title,titletoc,header]{appendix}

\begin{document}
\author{S. J. Cooper}
\email{scooper@star.sr.bham.ac.uk}
\affiliation{Institute for Gravitational Wave Astronomy University of Birmingham, Birmingham, B15 2TT, UK}
\author{C. J. Collins}
\affiliation{Department of Physics, University of Maryland, College Park, MD 20742, USA}
\author{L. Prokhorov}
\affiliation{Institute for Gravitational Wave Astronomy University of Birmingham, Birmingham, B15 2TT, UK}
\author{J. Warner}
\affiliation{LIGO Hanford Observatory, Richland, Washington 99352, United States of
America}
\affiliation{LIGO Laboratory, California Institute of Technology, Pasadena, CA 91125, United
States of America}
\author{D. Hoyland}
\affiliation{Institute for Gravitational Wave Astronomy University of Birmingham, Birmingham, B15 2TT, UK}
\author{C. M. Mow-Lowry}
\affiliation{Vrije Universiteit Amsterdam, 1081 HV, Amsterdam, Netherlands}

\title{Interferometric sensing of a commercial geophone}
\date{\today}

\begin{abstract}
We present a modified commercial L-4C geophone with interferometric readout that demonstrated a resolution 60 times lower than the included coil-magnet readout at low frequencies. The intended application for the modified sensor is in vibration isolation platforms that require improved performance at frequencies lower than 1\,Hz. A controls- and noise-model of an Advanced LIGO `HAM-ISI' vibration isolation system was developed, and it shows that our sensor can reduce the residual vibration by a factor of 70 at 0.1\,Hz. 
\end{abstract}
\maketitle

\section{Introduction}
With the completion of its first three observing runs the Advanced Laser Interferometer Gravitational-wave Observatory (LIGO) \cite{AdvancedLIGO15} has made incredible advancements in the field of astrophysics, with two of the highlights being the first direct detections of gravitational waves \cite{GW150914} and merging neutron stars heralding a new era of multi-messenger astronomy \cite{GW170817,GW170817MM}. 

To isolate the test masses from ground motion, LIGO uses a complex system of passive \cite{Aston12,Cumming2012} and active isolation \cite{Matichard15}. The active isolation is largely provided by the ISI (Internal Seismic Isolation), that reduces low-frequency motion and holds the in-vacuum optical benches at the correct position and orientation.
Despite the success of these systems, the lowest frequencies in the LIGO detection band, 10-20\,Hz, are limited by technical noise sources that are in turn caused by residual motion at even lower frequencies \cite{martynov2015thesis,Martynov16Sensitivity}. 

This residual motion must be reduced in order to detect gravitational waves at frequencies as low as 10\,Hz \cite{Yu185Hz} 

The ISIs are sensed by a mixture of displacement sensors, geophones, and force feedback seismometers. The instrument outputs are `blended' or `fused' together to use the best instrument in each frequency band. Two different models of commercial geophones, the Sercel L-4C and Geotech GS-13, are used throughout the detector to measure both the ground motion and the motion of the isolated platforms. While the performance of force feedback seismometers (Nanometrics T240 and Streckeisen STS-2) is inherently superior to that of geophones at frequencies below 0.5\,Hz, they are considerably more expensive. As such, only the most critical isolation platforms that house the primary test-masses and beamsplitter use broadband seismometers. Other chambers that house the auxiliary optics must rely on geophones to provide inertial platform measurements.

Coil-magnet geophones are limited by their intrinsic readout noise, caused by the Johnson noise in the geophone coils \cite{Kirchhoff2017}. The other intrinsic noise source, suspension thermal noise, is around a factor of 200 lower at 10\,mHz. With a sufficiently low-noise readout mechanism, the resolution of these sensors can be substantially improved at low frequencies. In the past interferometers have been used to decrease the readout noise contribution in broadband seismometers \cite{Zumberge10,Watchi2016} and thus increase the resolution of these devices. In an earlier paper we built a displacement sensor with self-noise low enough to measure suspension thermal noise from 10\,mHz to 2\,Hz \cite{cooper2018compact}. 

Improvements to the resolution of geophones will in turn allow for a re-design of the seismic isolation system's blending filters and control loops, substantially reducing sensor noise injection. In particular, they will allow for much improved inertial isolation between 0.1 and 0.3\,Hz and reduce the velocity of the platform at critical frequencies. 

In this paper we present a fibre-coupled interferometrically sensed L4-C that operates close to suspension thermal noise limit and has 60 times lower self-noise at 10\,mHz than the unmodified instrument.

\section{Noise sources}

Geophones are devices that measure seismic motion, they are comprised of a sensor coil surrounding a permanent magnet. Either the coil or magnet is attached to a suspended mass, while the other is attached to the device’s housing. At frequencies below their fundamental mechanical resonance, they are increasingly insensitive due to the reduction in relative motion between the magnet and coil. This causes the measured signal to decay proportional to frequency squared below the device's resonant frequency, while many of the noise terms remain roughly constant or increase at low frequencies.

To recover the input ground motion below resonance, the signal must be multiplied (in the frequency domain) by the inverse of the geophone's mechanical response, a process often called plant inversion, the inverse response $P^{-1}$ is given by the equation,
\begin{eqnarray}
P^{{-1}} & = & \frac{-\omega^2 + \frac{i\omega \omega_0}{Q} + \omega^2_0 }{\omega^2}. \label{eqn:l4cResp}
\end{eqnarray}

Where $\omega$ is the angular frequency, and $\omega_0$ and $Q$ are the (angular) resonant frequency and quality factor of the proof-mass' mechanical resonance. Plant-inversion correctly re-shapes the signal, but noise terms increase rapidly at low-frequencies compared to the measured signal, and the resulting output for typical levels of ground motion is dominated by self-noise below approximately 0.1\,Hz for the L-4C.  

The main noise sources for the interferometric L-4C are suspension thermal noise, readout noise, and laser frequency noise. The suspension thermal noise is calculated using the properties of the L-4C, this has a resonant frequency of 1\,Hz, a quality factor of 2.5 and a mass of 1\,kg. The equations given by Saulson \cite{Saulson90Thermal} are propagated through the plant, as shown in \cite{Harms_2017}. The thermal noise is then given by the equation, 

\begin{eqnarray}
	X_{g,th} = \sqrt{4k_BTm\omega_{0}^2\frac{\phi}{\omega}}\frac{1}{m\omega^2},
\end{eqnarray}
where $X_{g,th}$ is the thermal noise, $m$ is the mass, T is the temperature and $\omega_0$ is the resonance frequency of the inertial sensor. \\

Another source of readout noise is the laser frequency noise, and can be calculated using the equation,  
\begin{eqnarray}
\delta L = \frac{L \delta f}{f_0},
\end{eqnarray}
where $L$ is the length mismatch between the two arms, $f_0$ is the (optical) frequency of the laser and $\delta f$ is the laser frequency noise. To minimise the input laser frequency fluctuations we used a 1064\,nm solid-state Innolight Mephisto 500NE laser which has frequency noise of $\sim10^{5} \frac{Hz}{\sqrt{Hz}}$ at 0.1\,Hz and our interferometer had an arm length mismatch of 1\,mm.

The readout noise trace is based on the sensitivity curve presented in \cite{cooper2018compact}, which was measured with an smaller effective path length difference of 0.7\,mm. 
Both the laser frequency noise and readout noise are filtered by the plant-inversion filter of the L-4C from equation \ref{eqn:l4cResp}.

\begin{figure}[h!]
	\centering
	\includegraphics[trim = {1.5cm 0cm 1.5cm 0cm},clip, width=1\linewidth]{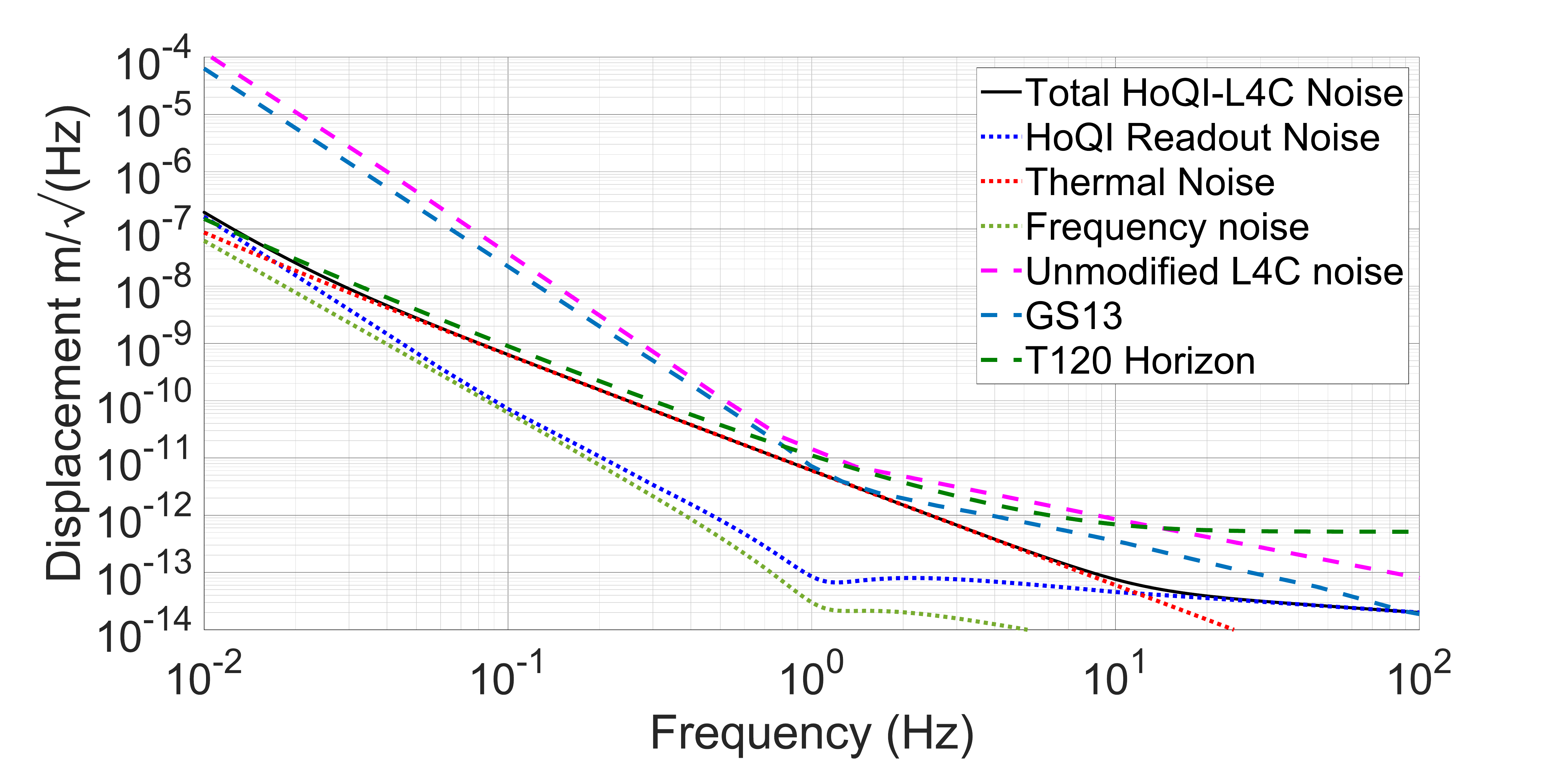}
	\caption{A noise budget of the optical inertial sensor. The interferometer readout noise (blue), structural thermal noise (red) and frequency noise (yellow) are summed in quadrature to produce the expected resolution of the optical L-4C (black). For comparison the resolution of an L-4C with coil readout is shown in magenta. }
	\label{fig:optNoiseBudget}
\end{figure}

A noise budget highlighting the dominant noise sources is shown in figure \ref{fig:optNoiseBudget}.  

Below 60\,mHz the device will be limited by frequency noise coupling, as this has a $f^{-3}$ dependency below the resonant frequency of the L-4C. In future instruments this can be reduced by either minimising the length mismatch between the two arms or stabilizing the laser frequency. At 10\,mHz frequency noise is only a factor of two higher than the suspension thermal noise, so only minimal changes are required in our chosen sensitivity band. Suspension thermal noise limits the resolution of the device between 60\,mHz and 2\,Hz. At frequencies above 2\,Hz frequency noise and electronic noise are the only significant sources of noise.  

\section{Construction of Devices}

To measure the self noise of interferometric L-4Cs the sensors must be isolated from external sources of motion, such as air currents, vibrations, thermal effects, and acoustic noise. The largest of these is seismic motion that is on the order of 1\,$\mu {\rm m}/\sqrt{\rm Hz}$ at approximately 0.15\,Hz. While this motion is large, it is coherent between multiple sensors provided that they are placed close together. Kirchhoff \cite{Kirchhoff2017} showed that the self-noise of a group of sensors could be measured in the presence of a large background signal by performing a huddle test whereby the coherent ground motion is subtracted from the a reference sensor to reveal its self-noise. 
This technique requires the sensors to have large dynamic range and be close together in order to subtract as much of the large background seismic signal as possible, while still having the precision required to sense the underlying sensor self-noise. 

The outer can of each L4C was removed to gain access to the moving proof mass and to prepare the outer can for the attachment of HoQI. Mounting holes were tapped and a large hole in the outer can was milled to allow the attachment of a light-weighted mirror mount to readout the motion of the proof mass. 

The interferometer used is an upgraded version of the sensor presented in \cite{cooper2018compact}, with the baseplate measuring 8\,x\,6\,cm that was rigidly attached to the outer can of the L4-C using three M2 bolts. Improving the coherence between multiple interferometric L4-Cs drove several design changes. We used a more stable fibre coupler, and improved waveplate and beamsplitter mounting structure resulting in approximately a factor 10 more coherence between adjacent sensors. Further details can be found in \cite{Cooper2019PhD}. An image of the prototype device is shown in figure \ref{fig:hoqiL4CV3} 

\begin{figure}[!h]
	\centering
	\begin{minipage}{.5\linewidth}
	  \centering
	  \includegraphics[width=0.9\textwidth]{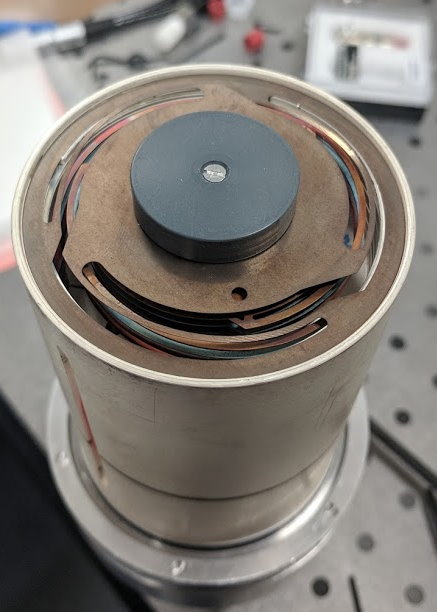}
	  \label{fig:exposedL4C}
	\end{minipage}%
	\begin{minipage}{.5\linewidth}
	  \centering
	  \includegraphics[width=0.9\textwidth]{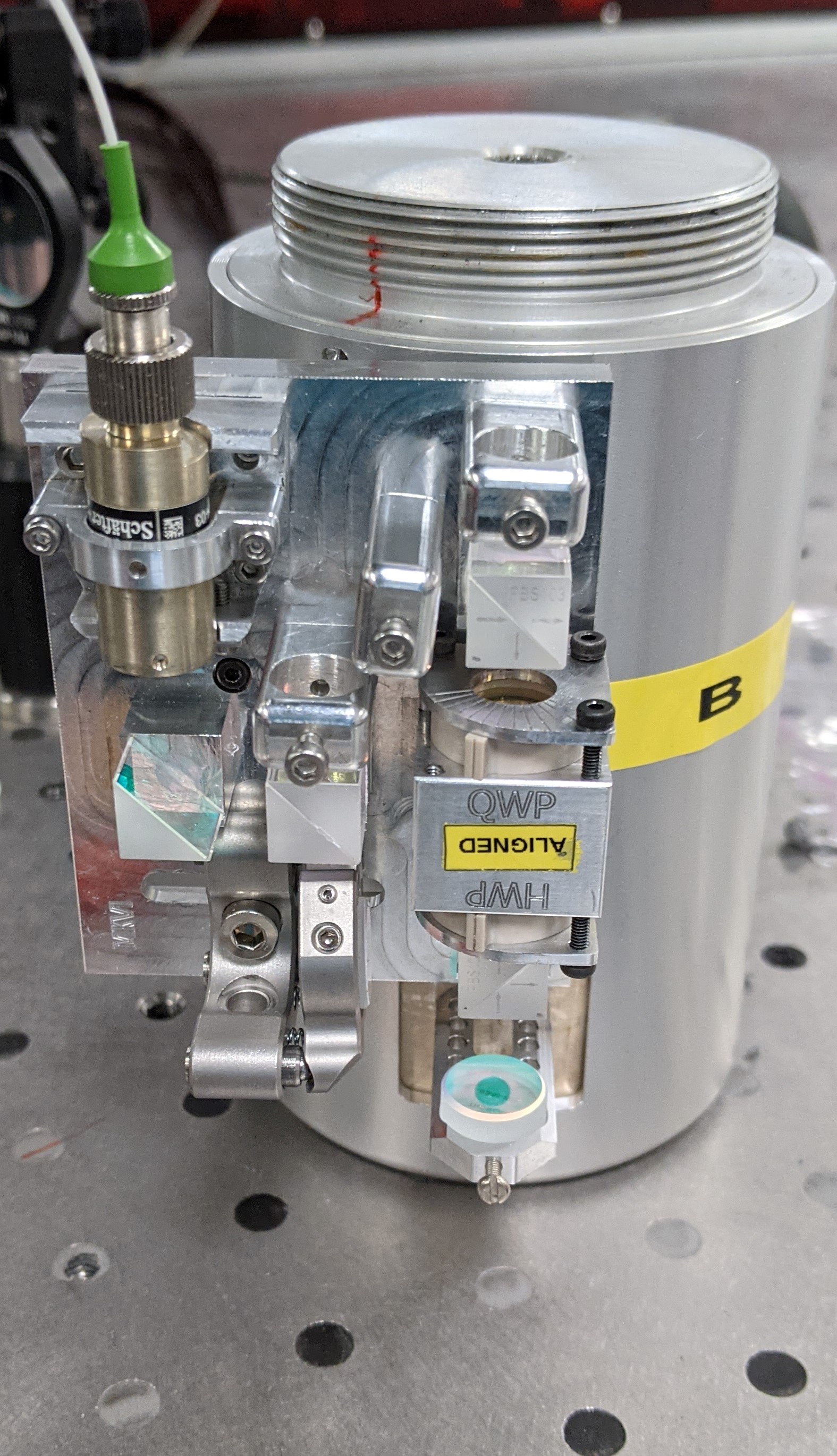}
	  \label{fig:hoqiL4CV3}
	\end{minipage}
	  \caption{a) Image showing the internal leaf springs of the L-4C during modification. b) Prototype version of HoQI used as a readout for an inertial sensor. The baseplate measures 8\,x\,6\,cm and is fixed to the side of an L-4C geophone.}
	\end{figure}
The length of the reference arm was adjusted to match the length of the readout arm to better than 1\,mm when the moving mass is at rest, minimising frequency noise coupling. The compactness of the device and its side mounting allows us to continue use the geophone coil-readout in parallel, providing a linear, calibrated, and high-coherence verification signal during operation.

The input laser light is injected into the interferometer through a single-mode polarisation-maintaining fibre. The beam is steered by a 45 degree mirror before being transmitted through a polarising beamsplitter to maximise the purity of the interferometer input polarisation, improving on the previous design by reducing the amplitude of spurious interferometers and increasing fringe visibility. The input alignment is handled by a custom mount for the collimator which is adjusted such that the laser light strikes the test mirror and returns to the photodiodes. The reference beam is then aligned to overlap with the test beam using a half-inch vacuum-compatible kinematic mount, thereby maximizing the fringe visibility at the interferometer readout.

\begin{figure}[!h]
	\centering
	\includegraphics[width=0.7\linewidth]{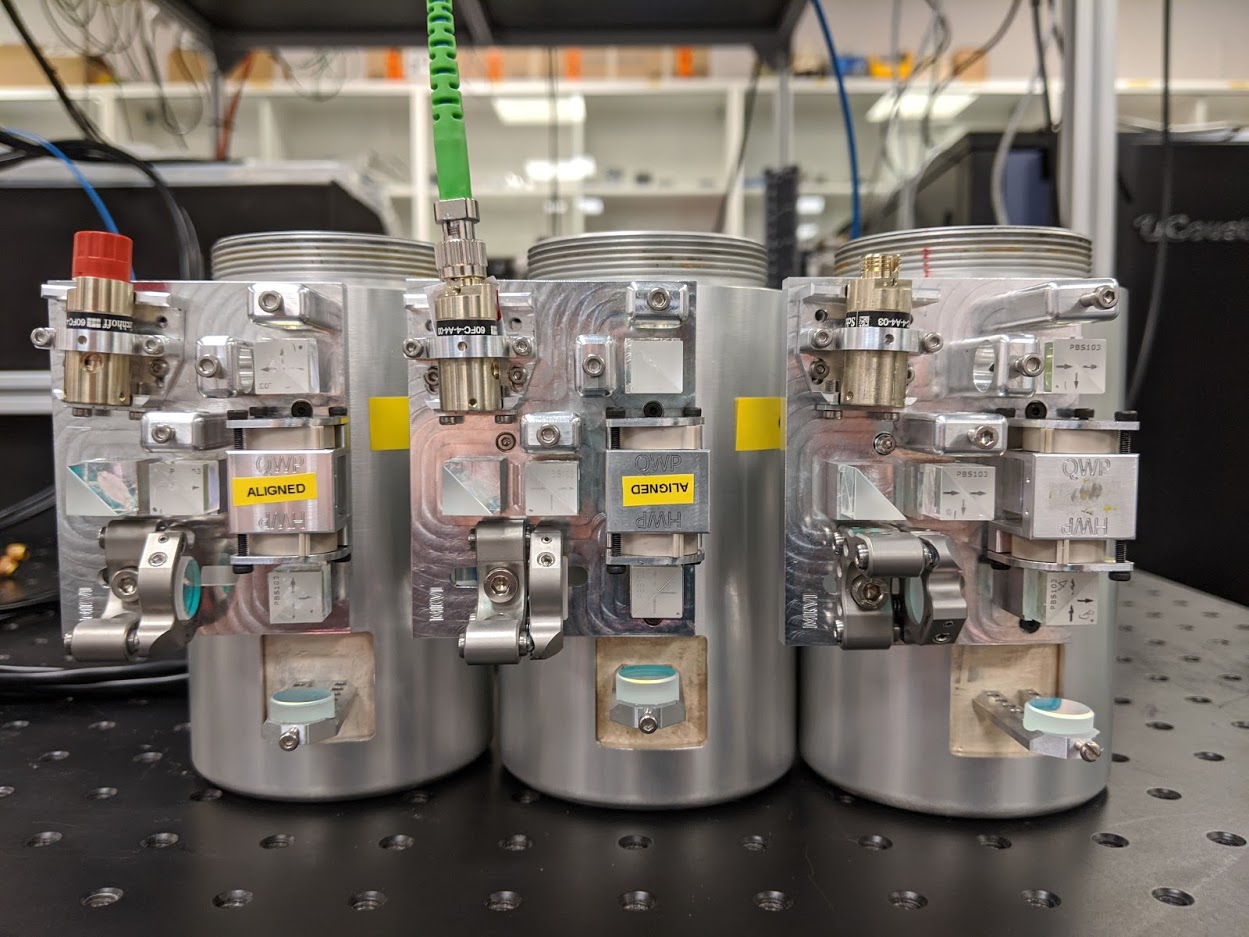}
	\caption{A photo of the three modified interferometric L-4Cs before they were huddle tested.}
	\label{fig:hoqiL4CV3}
\end{figure}

\begin{figure}[!h]
	\centering
	\includegraphics[width=0.9\linewidth]{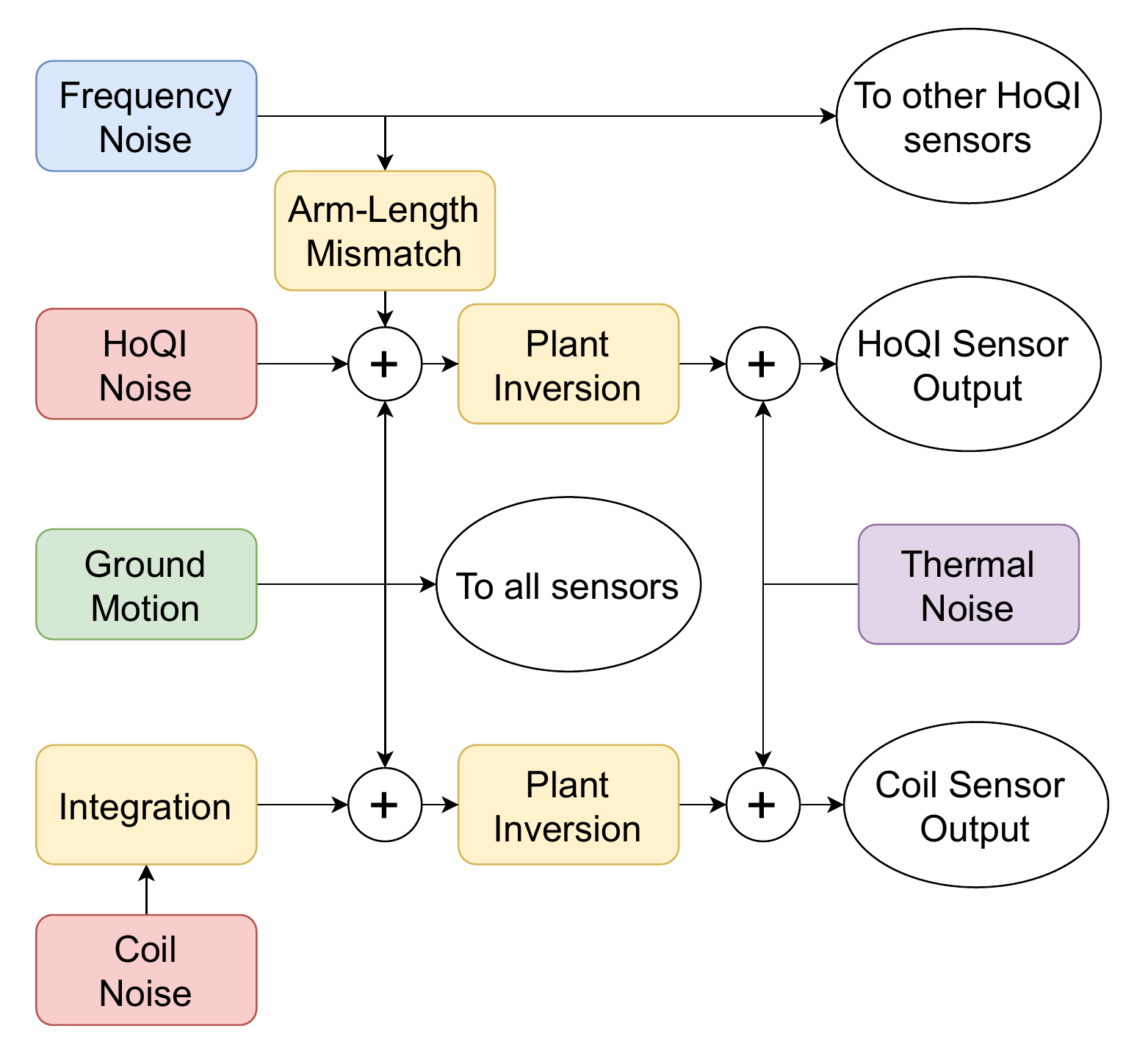}
	\caption{A block diagram illustrating different noise sources present in the device. Uncorrelated noises that cannot be suppressed using a huddle test are shown in red, correlated motion is shown in green. The frequency noise is common between all HoQI-L4C sensors, while thermal noise is common between the interferometric readouts and coil readouts on an individual sensor but are uncorrelated between multiple sensors.}
	\label{fig:blockdiagram}
\end{figure}

To determine the self-noise of the interferometric geophones, three L-4Cs utilising interferometric and coil readouts were placed close together and huddle tested inside an insulated box to minimise temperature, acoustic, and air current induced noise. The residual signal from an overnight measurement was processed using a MATLAB script (multi-channel coherence subtraction), developed by Brian Lantz and Wensheng Hua, removing any coherent signal present in the reference sensor(s) from the target sensor. This method is described in more detail by Kirchhoff \cite{Kirchhoff2017}. A block diagram illustrating the correlated signals which can be suppressed
using this technique, and uncorrelated noises which cannot be removed,
is shown in fig \ref{fig:blockdiagram}.

\section{Results}

\begin{figure}[!h]
	\centering
	\includegraphics[trim = {1.5cm 0cm 1.5cm 0cm},clip, width=1\linewidth]{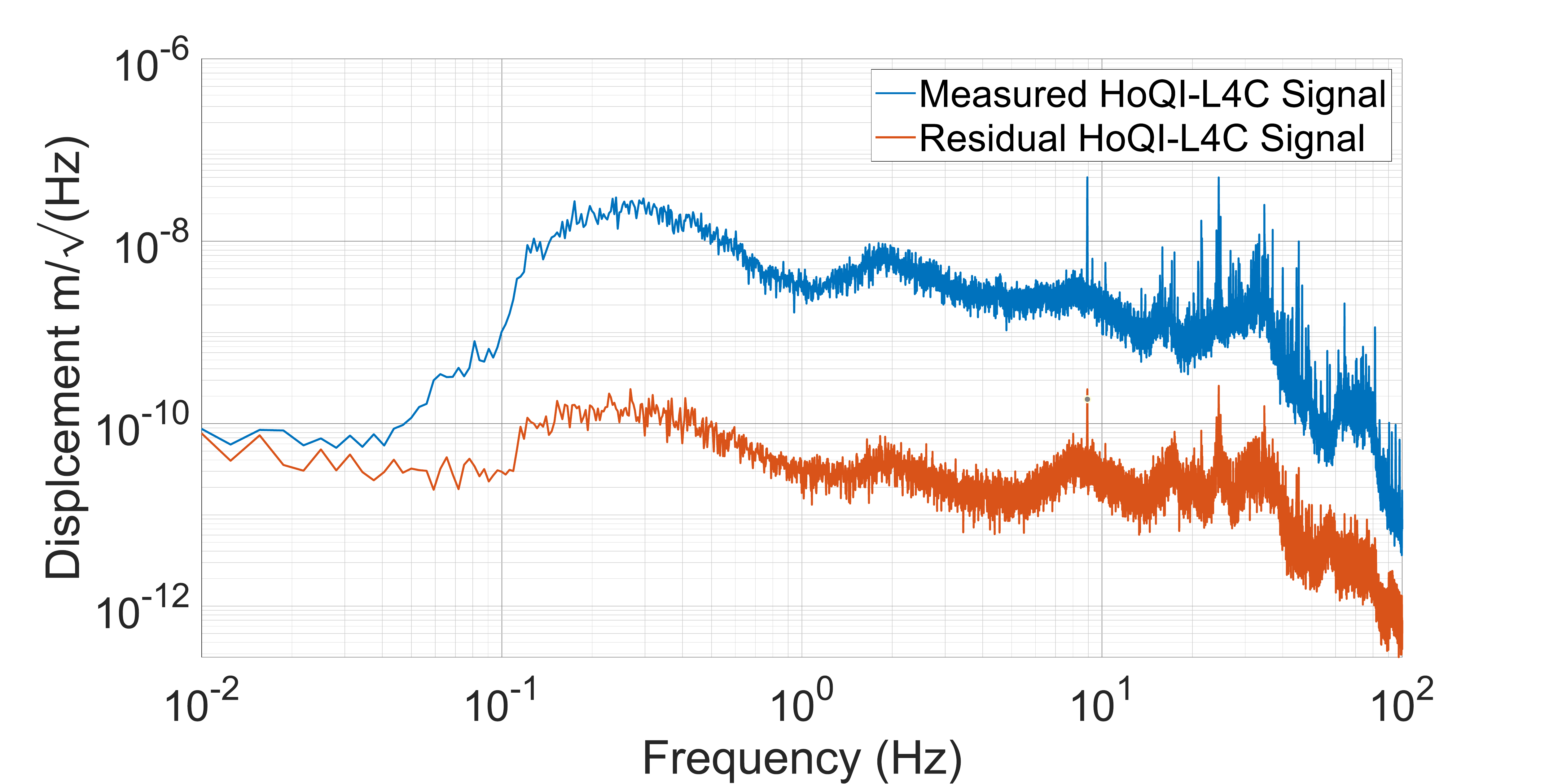}
	\caption{The spectral density of the measured signal from a HoQI-L4C (blue) and the residual spectrum after all coherent information from other sensors is removed (red). The mechanical response of the sensor has not been accounted for.}
	\label{fig:hoqicoilasd}
\end{figure}

Figure \ref{fig:hoqicoilasd} shows the raw output of one interferometer mounted on an L4-C when the device was placed on the laboratory floor. To extract the resolution of the HoQI-L4Cs in the presence of large seismic motion, three L-4Cs with optical and conventional readout were measured simultaneously. One interferometer signal was selected as the `target' and all coherent information from the remaining signals was removed to produce the `residual' signal trace. This residual signal is comprised of that interferometer's sensor noise and the uncorrelated components of input vibration, acoustic, and thermal effects. 

\begin{figure}[h!]
	\centering
	\includegraphics[trim = {1.5cm 0cm 1.5cm 0cm},clip, width=1\linewidth]{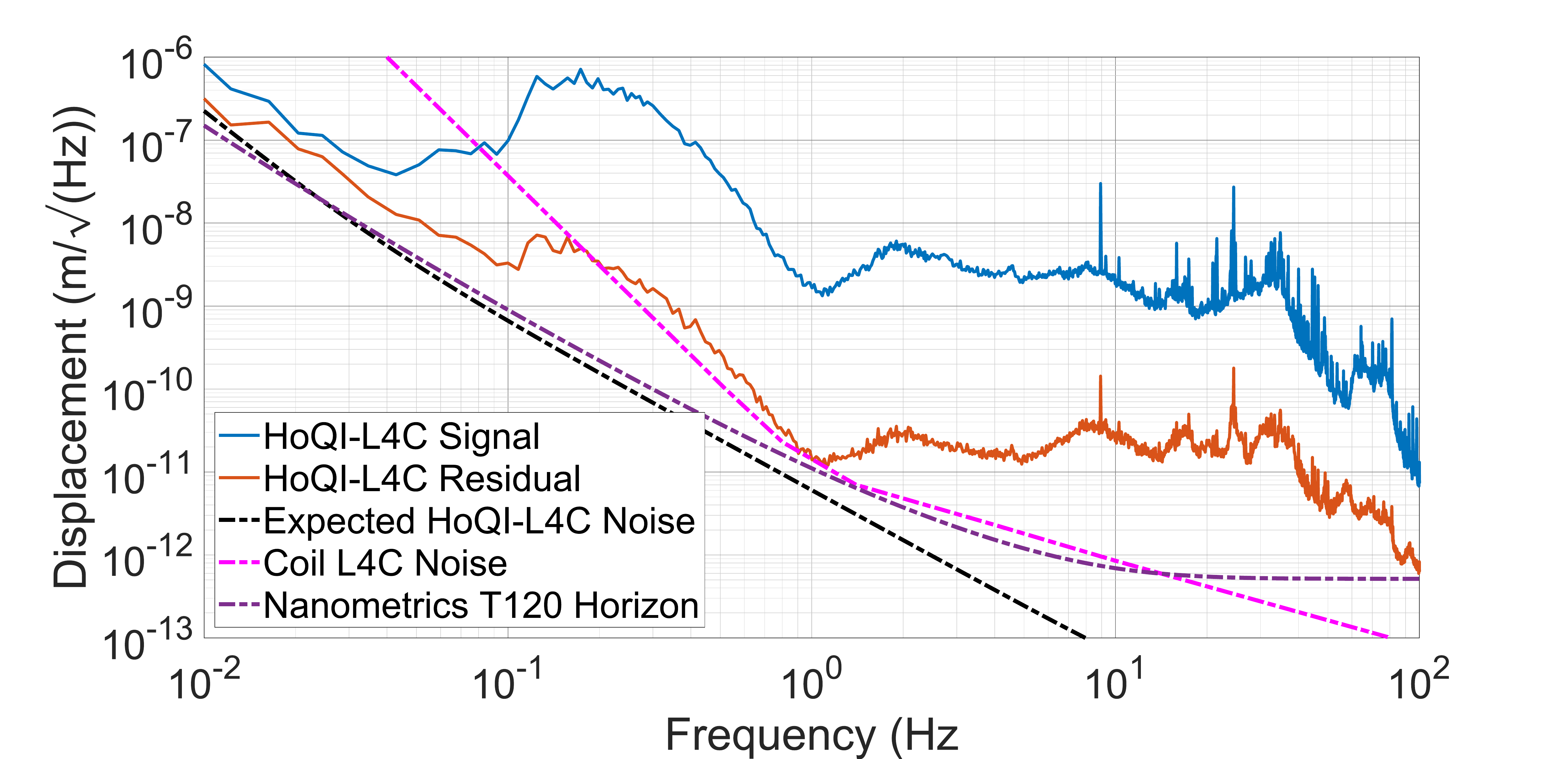}
	\caption{A comparison of the measured and predicted resolution of the HoQI-L4C with other commercial sensors.}
	\label{fig:hoqimeasuredpredicted}
\end{figure}

The measured signal and residual traces are compared with the calculated resolution of the HoQI-L4C sensor and two similar seismometers in figure \ref{fig:hoqimeasuredpredicted}. From 10 to 100\,mHz and at 1\,Hz the sensor resolution is within a factor of two of the predicted, thermal noise limited,  sensor resolution shown in dashed black. Between 0.1 and 1,Hz, the factor of 100 suppression of common motion is insufficient to suppress the large input motion caused by the microseismic peak. Above 1\,Hz insufficient coherence and vibration isolation prevents the residual motion from reaching the expected sensor resolution. 

A driving application for these sensors is in vibration isolation platforms used in gravitational wave detectors such as LIGO. To estimate the isolation performance improvement afforded by HoQI-L4Cs, a detailed numerical model of a HAM-ISI vibration isolation platform was constructed. The model itself is detailed in chapter 5 of \cite{Cooper2019PhD}. The numerical model reconstructs the closed loop performance of the HAM-ISI platform with good fidelity using only inputs from out-of-loop seismometers located near or on the platform’s support structures and knowledge of the plant response and loop gains. From these calculations, we can estimate the effects of sensor noise and feedback-control filters. 

We subsequently modified the model to include the projected HoQI-L4C self-noise in place of the existing GS-13 noise. To take full advantage of the improved sensor noise, new blend filters were generated such that the RMS velocity of the equation, 
\begin{equation}
\rm{vel}(f) = \rm{HP}\times\rm{ISNoise} + \rm{LP}\times\rm{DispNoise},
\end{equation} 
was minimised. Here HP, LP are the high- and low-pass blending filters and ISNoise and DispNoise are the respective spectral densities of the self-noises of the inertial and displacement sensors.

\begin{figure}[!h]
	\centering
	\includegraphics[trim = {1.5cm 0cm 1.5cm 0cm},clip, width=1\linewidth]{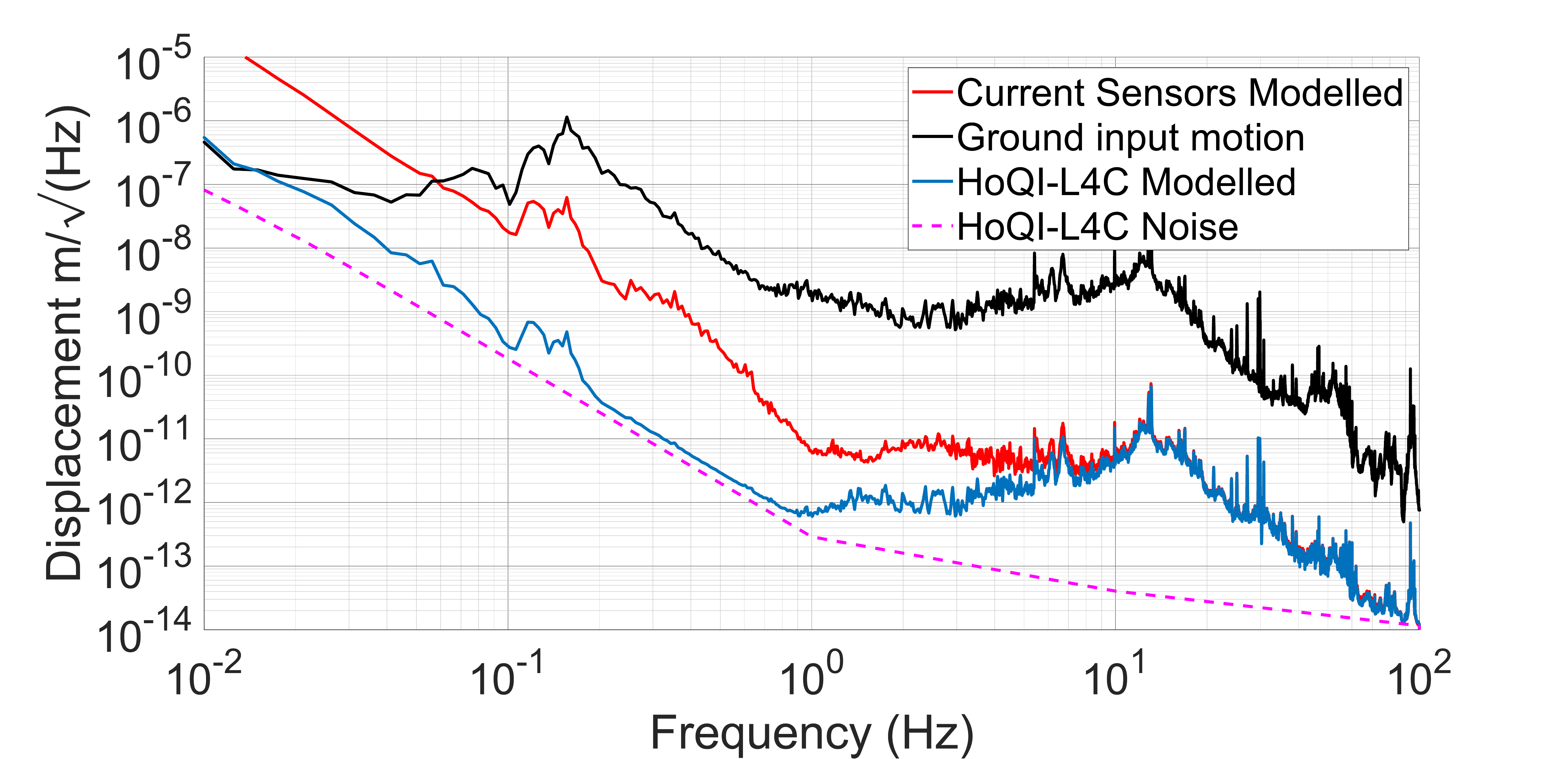}
	\caption{a comparison between the isolation performance using current sensors (red) and HoQI-L4Cs (blue). Inertial sensor noise and input ground motion are shown in dashed-magenta and black respectively.}
	\label{fig:ifoISI}
\end{figure}

The model was run using typical seismic input data, HoQI-L4C inertial sensors, and with the newly constructed blend filters. Existing feedback-control filters were not changed and we do not expect any change in the loop stability. Figure \ref{fig:ifoISI} shows a comparison between the estimated platform motion using existing sensors and the new configuration, with an estimated 70-fold reduction in motion at 0.1\,Hz and 10-fold reduction at 2\,Hz.

\section{Conclusion}
In this paper we have detailed the development of an L-4C seismometer that has been adapted to use a high resolution interferometric readout. These modifications increase the resolution of the sensor by factor of 60 at 10\,mHz and a factor of 5 at 10,Hz over the unmodified instrument. Through early testing, this resolution has been experimentally verified between 10 and 100\,mHz, however residual ground motion has prevented the modified device's resolution being currently verified at higher frequencies.numerically simulated model of an Advanced LIGO HAM-ISI was created to quantitatively predict the improvement that this could provide in suppression of ground motion, compared to vibration isolation systems currently used in LIGO. This model estimates that interferometrically readout L-4Cs can increase the isolation performance by a factor of 70 at 0.1\,Hz and a factor of 10 at 2\,Hz. 

\begin{acknowledgments}
	We thank John Bryant for technical support, Maura Shea and Isaac Davis for integration and testing of the data acquisition systems used in this project. The authors acknowledge the support of the Institute for Gravitational Wave Astronomy at the University of Birmingham, STFC 'Astrophysics at the University of Birmingham' grant ST/S000305/1. 
\end{acknowledgments}

\end{document}